\documentstyle[epsf,rotate]{mn}       

\begin{document}

\title[A TiO study of the dwarf nova IP Pegasi]
{A TiO study of the dwarf nova IP Pegasi}

\author[Beekman et al.]
{
G. Beekman, M. Somers, T. Naylor and C. Hellier\\
{\it Department of Physics, Keele University, Staffordshire ST5 5BG}
}

\date{Accepted ???.
      Received ???;
      In original form ???.}

\pagerange{\pageref{firstpage}--\pageref{lastpage}}

\pubyear{}

\maketitle

\label{firstpage}

\begin{abstract}
We present red spectra in the region $\sim\lambda$7000--8300\AA\ 
of the eclipsing dwarf nova IP Peg, with simultaneous narrow-band
photometry centered at 7322\AA.
We show that by placing a second star on the slit we can correct 
for the telluric absorption bands which have hitherto made the TiO features 
from the secondary star unusable.
We use these TiO features to carry out a radial velocity study of the 
secondary star, and find this gives an improvement in signal-to-noise
of a factor two compared with using the Na{I} doublet. 
In contrast with previous results, we find no
apparent ellipticity in the radial velocity curve.
As a result we revise the semi-amplitude to $K_{2}=331.3\pm 5.8$ km s$^{-1}$,
and thus the primary and secondary star masses to 
$1.05^{+0.14}_{-0.07}$M$_{\odot}$ and 
$0.33^{+0.14}_{-0.05}$M$_{\odot}$ respectively. 
Although this is the lowest mass yet derived for the secondary star, it
is still over-massive for its observed spectral type.
However, the revised mass and radius bring IP Peg into line with other CVs
in the mass-radius-period relationships.

By fitting the phase resolved spectra around the TiO bands to a mean 
spectrum, we attempt to isolate the lightcurve of the secondary star.
The resulting lightcurve has marked deviations from the expected  
ellipsoidal shape.
The largest difference is at phase 0.5, and can be explained as an eclipse
of the secondary star by the disc, 
indicating that the disc is optically thick when viewed at high 
inclination angles. 

\end{abstract}

\vspace{5mm}

\begin{keywords}
novae, cataclysmic variables -- binaries: close  -- binaries: spectroscopic
-- binaries: eclipsing -- stars:fundamental parameters -- 
stars: individual: IP Peg
\end{keywords}


\section{Introduction} 

IP Peg is a dwarf nova which 
brightens by $\sim$2 mag every 100 days or so, with each outburst lasting
$\sim$2 weeks, and is important because it is the brightest known eclipsing
dwarf nova above the period gap.
The eclipses give tight constraints on the geometry of the system, and
in combination with other data, such as radial velocity studies, allow 
the masses of the two stars to be determined.
We should thus be able to compare the mass, radius and density of the
secondary star with those of main sequence stars, and learn something of its
structure and evolution.
Unfortunately the radial velocity curve of the secondary star in IP Peg is 
problematical, being apparently elliptical (Martin et al. 1989, 
henceforth M89).
The star is believed to be in a circular orbit, and the apparent ellipticity 
is thought to be due to irradiation of the 
secondary star (although we shall present an alternative explanation in 
Section \ref{ecc_dis}), and an uncertain
correction must be applied before the binary parameters can be determined.
The result is acutely embarrassing.
Smith \& Dhillon (1998) collect together all the available masses, radii and
spectral types for CVs.
IP Peg is one of the few objects that does not rely on the dubious method
of using the accretion disc lines to determine the white dwarf radial
velocity, and so should give reliable parameters.
However, in all the relationships IP Peg is a persistent offender, lying well 
clear of the supposedly less reliable points in both the mass vs. orbital 
period and mass vs. spectral type diagrams.

A further complication is presented by the lightcurve of the system.
High-speed photometric observations show a light curve dominated by a
bright spot and a deep eclipse. The eclipse is that of the bright spot
and white dwarf by the secondary star, superposed on a
more gradual disc eclipse. 
Unfortunately the  bright spot ingress overlaps 
with that of the white dwarf, making the ephemeris determination
more problematical. 
Wood \& Crawford (1986) were able to use their high-speed photometric
observations to derive an ephemeris for IP Peg based on white dwarf egress 
timings. 
Later observations by Wood et al. (1989) and Wolf et al. (1993) show that 
the white dwarf egress varies markedly in strength and duration
(10s to 300s).

In this paper we present red spectroscopy and photometry of IP Peg.
The observations and their reduction are explained in Section \ref{obs}.
After that the paper divides into two main threads.
The first is the radial velocity study and its results.
The extraction of the velocities is described in Section \ref{rad_vel} 
after which we discuss the summed spectra and new ephemeris (Sections
\ref{disentangle} and \ref{ephemeris}).
In Section \ref{circularity} we discuss the absence of apparent eccentricity 
in our data, and conclude the system has changed in some way.
We use our radial velocity semi-amplitude to derive new system parameters
in Section \ref{masses}.
The second thread, isolating the lightcurve of the secondary star, is
explored in Section \ref{flux_deficits}.


\section{Observations}

\label{obs}

Photometry and spectroscopy of IP Peg were obtained over four
nights in 1995 October from the 2.5-m INT and 1.0-m JKT on La Palma. Only
two nights of photometry were taken -- simultaneous with
the spectroscopy (see Section \ref{spec_phot}). Table \ref{observations}
contains a log of the observations. 

\begin{table*}
 \begin{minipage}{100mm}
  \caption{Journal of observations of IP Pegasi}
  \label{observations}
  \begin{tabular}{c c c c c c} 

   Date      & No. of & Wavelength  & Exposure   & Phase     & Mean Seeing \\
 (Oct. '95)  & Frames & Region (\AA)& Time (sec) & Covered   & (arcsec)    \\ \\
 Photometry: \\
     4th     &    81  & {\sc Oii} filter  & 120       & 0.51--1.82 & 2.17  \\
     5th     &   100  & {\sc Oii} filter  & 120       & 0.31--1.70 & 1.65  \\
 Spectroscopy: \\
     4th     &    18  & 7100--8300  & 600       & 0.38--1.03 & 1.85 \\
     5th     &    18  & 7100--8300  & 600       & 0.38--1.70 & 1.42 \\
     9th     &    22  & 7000--7800  & 600       & 0.56--1.72 & 1.78 \\
    10th     &    39  & 7000--7800  & 400       & 0.83--1.79 & 1.98 \\
  \end{tabular}
 \end{minipage}
\end{table*}

Data from the AAVSO show that in the latter half of
1995 IP Peg underwent two outbursts, one in mid-September and one in
mid-December. The data presented in this paper were taken in the first
week of October, approximately two weeks after the end of the 
mid-September outburst, and are thus during quiescence.


\subsection{Photometry}

\label{phot}

Data were taken with the JKT on 1995 October 4th \& 5th using an
{\sc eev7}-chip with a scale of 0.31 arcsec per pixel. A narrow band
({\sc fwhm}=46\AA) O{\sc ii} filter with maximum transmission at a 
wavelength of $\lambda$7322\AA \  was used on both nights. 

\begin{figure}
{\includegraphics{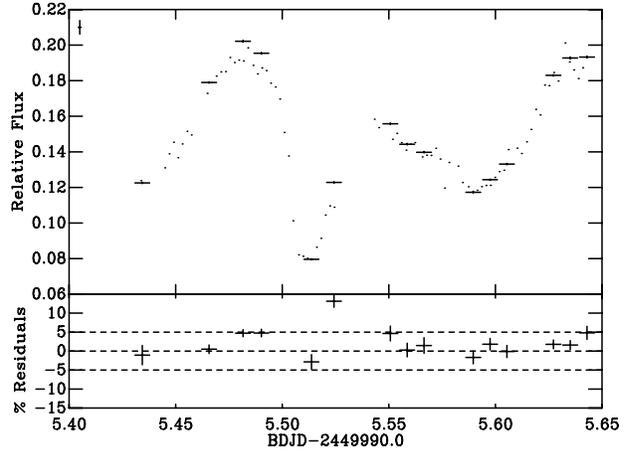}}
\vspace{2.5in}
\caption{
The lightcurves obtained from the photometry (dots) and spectroscopy
(bars).
The lower panel shows the percentage residuals between the two, calculated as 
$ 100.0*(R_{\rm spec}-R_{\rm phot})/R_{\rm phot} $.
The point in the top left of the upper panel represents the average
size of a photometric error. Data taken on 1995 October 4th. 
}
\label{4th}
\end{figure}

\begin{figure}
{\includegraphics{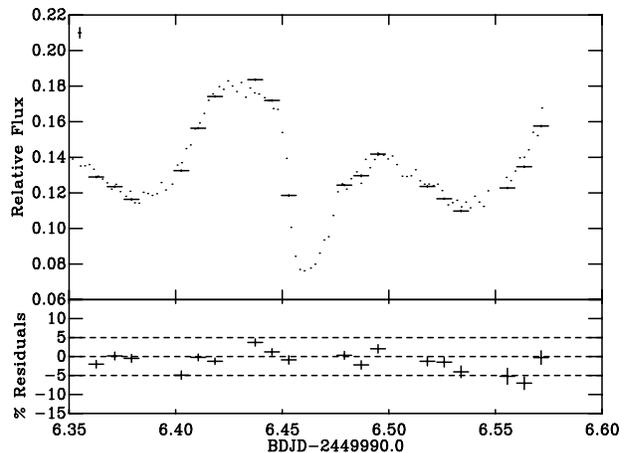}}
\vspace{2.5in}
\caption{As previous figure but for the data taken on 1995 October 5th.}
\label{5th}
\end{figure}

The images were processed by subtracting off a single bias image in the
case of the Oct. 4th data, and a median of three biases for the Oct 5th 
data.
A single sky flat taken at the beginning of each night was used
to flatfield the images. The photometry was then extracted using an
optimal extraction method (Naylor 1998). The photometry of IP Peg
was divided by that of a neighbouring star (the second star placed on the
slit during the spectroscopic run---see below) to remove the
effects of sky transparency variations and, thus, the photometry is relative. 
The phase-folded light curves for both nights are shown in Figs. \ref{4th} and
\ref{5th}
with the bright-spot hump at $\phi\sim -0.2$ and the primary eclipse 
at $\phi=0.0$. 

\subsection{Spectroscopy}

Data were taken on 1995 October 4th, 5th, 9th and 10th with the
INT, using a {\sc tek4}-chip with a scale of 0.33 arcsec per pixel,
but binned by a factor two in the spatial direction.
A two-arcsec slit was used set at PA 58.7 degrees to obtain the spectrum of 
a neighbouring star simultaneously (the star used to divide the IP Peg 
photometry by---see above). 
This allowed us to correct for slit losses (see Section \ref{spec_phot}).
The R831R grating was used on the 4th and 5th, giving $\sim$1.22
\AA\ per pixel; the R1200R grating was used on the 9th and 10th, giving
$\sim$0.84 \AA\ per pixel.  An arc spectrum was taken each time the telescope's 
position on the sky was changed ($\sim$ every hour). The nights of the 4th and
5th were generally clear but the nights of the 9th and 10th were affected by 
cloud, especially so the data of the 9th. An exposure time of 600s was used on
the 4th, 5th and 9th: 400s on the 10th.

The images for each night were first bias subtracted. 
Tungsten flats taken each night were then used, after dividing
by fitted low-order polynomials, to flatfield the images. Extraction of
the spectra was performed using an optimal extraction method (Horne 1986,
Robertson 1986), and then wavelength calibration was done using low-order 
polynomial fits to the appropriate arc spectra.


\subsection{Removing the effects of Earth's atmosphere}

\label{telluric}

\begin{figure*}
\includegraphics{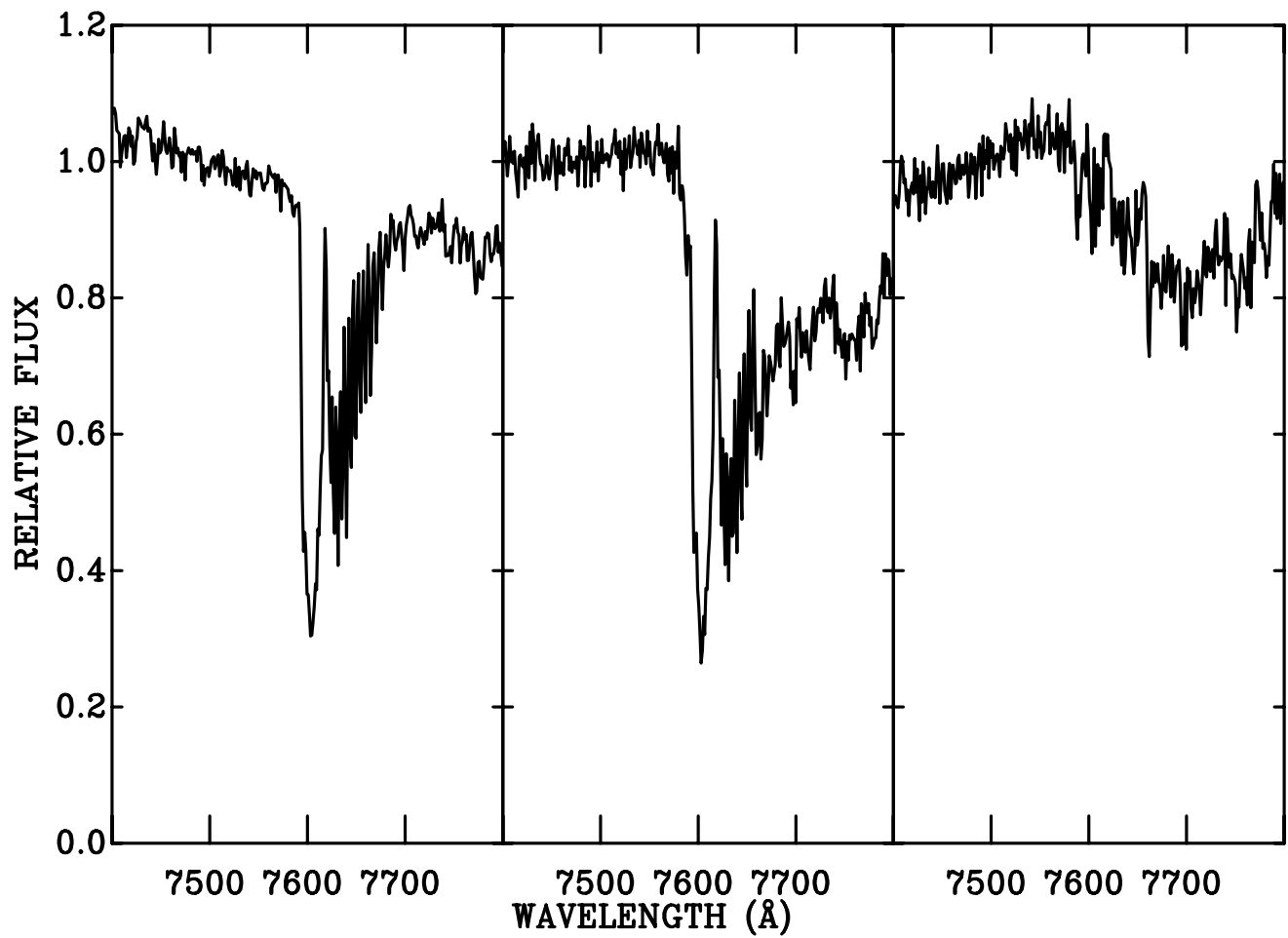}
\vspace{5.4in}
\caption{ 
This figure shows the correction for the atmospheric `A'-band near 
$\lambda$7600\AA.
The left hand panel is the second star on the slit; the center panel
is the uncorrected IP Peg spectrum; the right hand panel after correction. 
All the spectra have been normalised to one at 7500\AA.
Note how the drop of the TiO band heads (7550-7650\AA) and the KI lines
(7650 and 7700\AA) become visible after correction.
}
\label{A-band}
\end{figure*}

Normally, one observes a standard star (usually a near-featureless O star) 
at regular intervals throughout the night to divide the program star by. 
Here, we use the second star placed on the slit and observed simultaneously 
with IP Peg. 
This has the advantage that it avoids the assumption of uniformity of 
atmospheric absorption with position and time. 
Fig. \ref{A-band} shows that the atmospheric 
absorption has been well corrected, including the large bands near 7600\AA. 

\begin{figure}
{\includegraphics{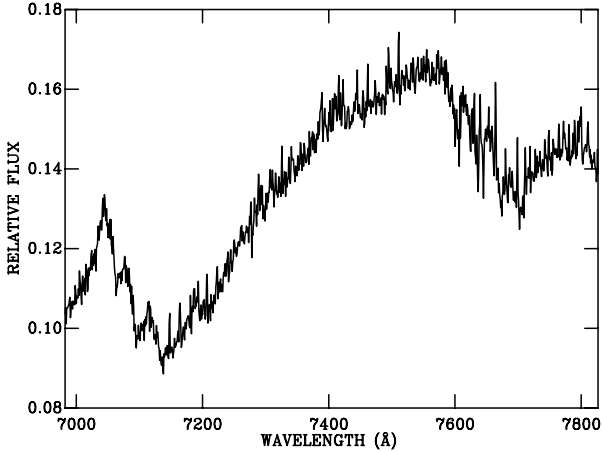}
}
\vspace{2.5in}
\caption{
The result of a simple mean of the spectra which 
result from dividing the IP Peg spectrum by the second star on the slit.
Only the higher-resolution spectra (1995 Oct 9 and 10) have been used.
}
\label{simple}
\end{figure}

The disadvantage of this technique is that the star
may not be as featureless as one would like, since its main selection 
criterion is proximity to the target star. 
Comparing blue spectra of the star with those in Jacoby, Hunter 
\& Christian (1984) shows it to be a late F-star.
Such stars are good atmospheric calibrators at low resolution (see
Mason et al 1995), but at the intermediate resolutions we used weak 
absorption features are present.
In Fig. \ref{simple} we have averaged all our higher-resolution divided 
spectra with no 
velocity shift, and can see these features as ``emission'' lines superimposed 
on a velocity smeared M-star spectrum
(since we have not corrected to the rest frame of the secondary star).
The features are weak (normally around a couple of percent, with a few
at 10 percent) and as we shall show in Sections \ref{rad_vel}
and \ref{disentangle}, have a negligible effect our radial velocity study and
final summed spectrum. 

\subsection{Relative spectrophotometry}

\label{spec_phot}

During our spectroscopic run with the INT, two nights of photometric data
(simultaneous with the spectroscopic data) were taken with the JKT
to test the two-star spectrophotometry technique. 
The wavelength region of the spectroscopy corresponding to the photometry 
was isolated and summed according to the weights of the O{\sc ii} filter.
After summing, the spectrophotometry of IP Peg was divided by the similarly 
summed flux from the second star on the slit. 

Figs. \ref{4th} and \ref{5th} show the photometric data over-plotted on the
spectroscopic data. 
The lower panels in the figures show the percentage residuals between the two,
calculated as $100.0*(R_{\rm spec}-R_{\rm phot})/R_{\rm phot}$.
Where more than one photometric point is coincident with a
spectroscopic one, a simple average of them is taken. 
For the spectroscopy point covering primary-eclipse egress in the data of 
1995 October 4th, the exposure was too large long sample the fast rate of 
change in brightness properly. 
This resulted in the relatively large residual seen here. 
Ignoring this point in the 
analysis, the average residual size for the nights of the 4th and 5th are 
3.28\% and 2.80\%, respectively. That is, the difference between relative 
photometry and relative spectrophotometry is only at the few percent level
and, for the most part, lies within the error of the photometry. 
Thus the real difference is probably well within the measurement errors. This
shows that performing relative spectroscopy by the inclusion of a second star 
on the slit works well.

This result is apparently in contradiction with the work of Webb et al.
(2000).
They performed a similar experiment to ours, obtaining spectroscopic
observations of the low-mass X-ray binary J0422+32, again with a second star
on the slit.
When they compared the relative photometry from their spectroscopic
observations with that from simultaneous photometry they found differences
of order 15 percent. 
The author in common between these two papers has no explanation as to why
this should be so, though we note that IP Peg is much brighter,
and that a broadband filter was used for J0422+32.


\section{A Radial Velocity Study} 
\label{rad_vel}

We used an iterative procedure to determine the radial velocity curve of 
IP Peg, in which a summed spectrum (created after each iteration) is used 
as the template. 
This technique is similar to that used by Mukai \& Charles (1988) and
Horne, Welsh \& Wade (1993).
First, the mid-white dwarf egress timings of Wood et al. (1989) and 
Wolf et al. (1993) were fitted with a linear ephemeris, and this ephemeris 
was initially used to phase all of our spectra. 
The radial velocity semi-amplitude ($K_{2}$) value from M89
was then used to assign a stellar radial velocity component 
to each spectrum for a given phase $\phi$ by assuming a circular orbit for 
the red star of the form $K_{2}\sin(2\pi\phi_{i})$, where $\phi_{i}$ is the 
phase of spectrum $i$.
Each spectrum was then Doppler-shifted by its orbital radial velocity; 
all such shifted spectra were then co-added into a grand-sum
spectrum.  
This grand-sum spectrum (less the spectrum of interest --- see later) was 
then used as a template spectrum against which to cross-correlate all the 
individual spectra to obtain a radial velocity study. 

The velocity extraction process is run iteratively. The velocities from the 
initial set of cross-correlations are fitted with a sine wave to return a new 
$K_{2}$-value and correction to phase zero; each individual spectrum is then 
assigned a new phase and new radial
velocity from this second generation ephemeris and $K_{2}$-value; each
spectrum is then shifted by these new velocities and combined
to make a new orbit-averaged template spectrum; the spectra are then
cross-correlated against this new average, new velocities extracted and so
on, and the whole process repeats until convergence. 

The process 
was considered to have converged when the correction to phase zero was less 
than 0.001 of the orbital period, corresponding to a correction of less than 
14 seconds. This value was chosen because tests showed that beyond this,
no significant improvement within the errors of the fit was achieved.
Convergence typically occurred within ten iterations. A small number of 
spectra returned dubious velocities; these spectra were of lower 
signal-to-noise (mainly those of 1995 October 9th, due to occasional
cloud), and were removed from the analysis.

Note that to avoid any danger of `self' cross-correlation, the spectrum
being correlated is not used in creating the average-spectrum template
against which it is being cross-correlated. 
This results in a new template
being created for each individual spectrum, but the large number of spectra
which adequately sample the orbit ensures that the template 
spectrum is, to all intents and purposes, identical throughout. 

The first two nights of data
were taken at a slightly lower resolution than the last two, so the data were
split into high and low resolution data sets and each one treated separately. 
For both sets of data, two regions were used in the cross-correlation but
only one region was used at a time. These regions were a 
115\AA-wide region from $\lambda$7105--7220\AA, and a 170\AA--wide region from 
$\lambda$7560--7730\AA. 
The slightly different wavelength coverage between the two data sets allowed 
the shorter wavelength cross-correlation region in the slightly higher 
resolution data set to be extended blue-wards to $\lambda$7000\AA.

\begin{table*}
 \begin{minipage}{140mm}
 \caption{ Radial velocity data using an orbital-average as template}
 \label{ecc}
 \begin{tabular}{c c c c c c c c c c c}

  $\lambda$ & Notes     & $N$ & $K_{2}$ & T$_{0} $-2449998.0$^2$ & $\sigma_{c}$ & $\sigma_{e}$ & $F$ & $p^{1}$\\
  7105--7220 & TiO       & 31  & 353.8 $\pm$ 14.1 & 0.51609(86) & 32.14 & 31.67 & 0.404 & 0.671 \\
  7560--7730 & TiO,K     & 32  & 325.6 $\pm$ 14.3 & 0.51722(95) & 33.92 & 33.41 & 0.424 & 0.658 \\
  7000--7220 & TiO       & 59  & 310.6 $\pm$ 9.4  & 0.51841(73) & 32.42 & 31.78 & 1.183 & 0.314 \\
  7560--7730 & TiO,K     & 59  & 344.3 $\pm$ 9.5  & 0.51612(64) & 33.01 & 32.88 & 0.220 & 0.803 \\
            & $w\Sigma$ &     & 331.3 $\pm$ 5.8  & 0.51692(39)    \\
\\
  7670--8319 & M89 region & 32 & 327.7 $\pm$ 16.1 &             & 30.11 & 30.11 & 0.00  & 1.00  \\
 \end{tabular}
 \medskip \\ $^{1}$Orbit is consistent with being circular for $p \geq 0.05$.
 \\ $^{2}$Numbers in brackets are the errors on the same number of figures 
 immediately to their left.
 \end{minipage}
\end{table*}

\begin{figure*}
\includegraphics{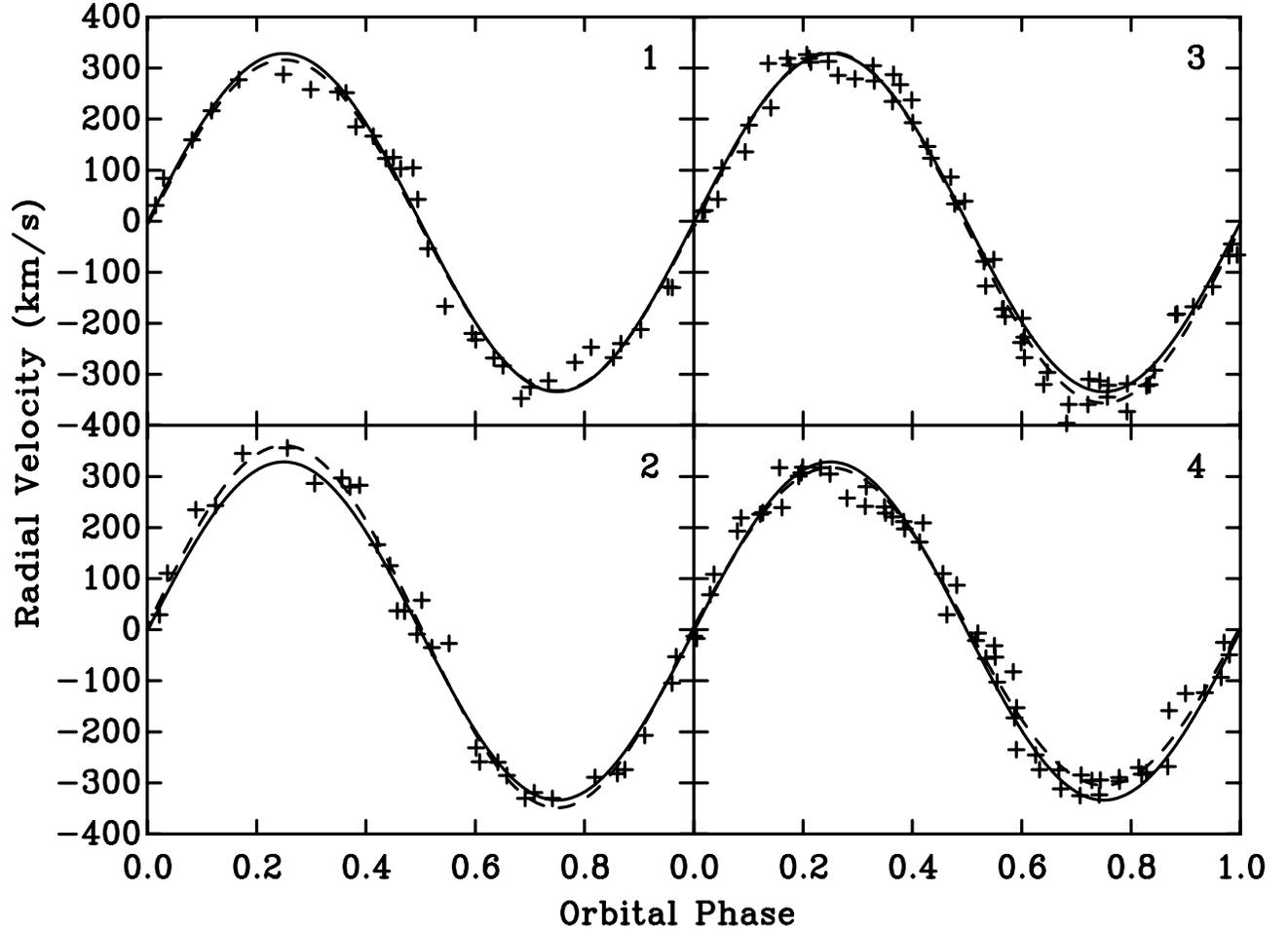}
\vspace{5.4in}
\caption{ 
Extracted IP Peg radial velocity curves. 
(1) The region 7560--7330\AA\ for Oct. 4th \& 5th;
(2) the region 7105--7220\AA\ for Oct. 4th \& 5th;  
(3) the region 7560--7730\AA\ for Oct. 9th \& 10th;
(4) the region 7000--7220\AA\ for Oct. 9th \& 10th. 
The dashed lines are the best fit to each panel (see Table \ref{ecc}).
The full lines are the weighted mean of the individual fits.
}
\label{v_panel}
\end{figure*}

Table \ref{ecc} presents the results of this study for the two regions
used in each data set.
A weighted mean (labelled $w\Sigma$ in the table) of
these results provides us with a new mid-white dwarf eclipse ({\it not}
egress) timing and $K_2$.
The individual radial velocity curves are shown in Fig. \ref{v_panel} with a
circular orbit fit using the above weighted mean values over plotted as
a full line; the individual orbital fits of Table \ref{ecc} are
over-plotted as dashed lines.
The quoted error is the 68\% confidence interval (`1$\sigma$') calculated for
two free parameters (Lampton, Margon \& Bowyer 1976).
To derive this error the radial velocity points were given equal weight, and
a $\chi_{\nu} ^2$ assumed for the best fit.  

We compared the signal-to-noise we obtained using the TiO bands
with that from the more traditional Na{I} doublet.
The gain using the new technique is about a factor two in signal-to-noise
if the results from the two TiO bands are averaged.
We also tried cross-correlating each spectrum against an M-star template 
instead of the mean IP Peg spectrum.
There was an increase in noise for the fits to these data, presumably 
because the mean spectrum gives a better match to the spectral type than 
any template star.

Finally we wished to ascertain whether the weak F-star features introduced
by our technique for removing telluric absorption (see Section \ref{telluric})
had affected our values for the radial velocity.
Specifically, there was a concern that it would introduce a signal at
zero velocity shift, which might reduce the semi-amplitude.
During the run we took several observations of early-type stars to check our
correction technique.
We chose a pair of observations where an O-star corrected the observation
of the second star on our slit well, yielding a telluric absorption corrected
spectrum of this star.
We then divided our corrected spectra of IP Peg by this spectrum, thus
doubling the strength of the F-star features.
We then repeated our cross correlation analysis for Oct. 4 and 5.
The derived semi-amplitude changed by less than the error (+7 km s$^{-1}$ for
7105-7220\AA\ and -4 km s$^{-1}$ for 7560-7730\AA).
It should be noted that this is an over-estimate of how much the F-star
affects our study, since we will also have introduced noise, and poor
atmospheric correction, all of which might tend to decrease the
semi-amplitude.
Despite this, it is clear that the second star on the slit makes a negligible
contribution to the cross-correlation function, for reasons we explain
in the next Section.


\section{The summed spectrum}
\label{disentangle}

\begin{figure}
{\includegraphics{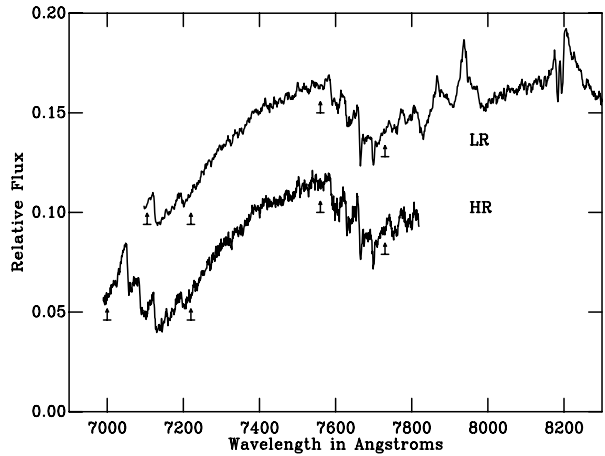}}
\vspace{2.5in}
\caption{ 
The Doppler-shifted, averaged spectra of IP Peg taken during 1995 October. 
LR is the lower resolution data taken on the nights of the 4th \& 5th 
(upper spectrum); HR is the higher resolution 
data taken on the 9th \& 10th (lower spectrum). Lower limit symbols mark the 
boundaries of the cross-correlation regions used in the radial velocity study.}
 \label{spectra}
\end{figure}

Fig. \ref{spectra} shows the final Doppler-shifted weighted-mean
spectrum.
The reasons for our choice of cross-correlation bands should now be clear.
The first corresponds to the TiO band heads around 7100\AA, the
second to the TiO band heads and K{I} lines ($\lambda$7764.9, 7699.0\AA) 
near 7600\AA.
The final feature obviously from the secondary star is the Na{I}
doublet ($\lambda$8183.3, 8194.8\AA).

\begin{figure}
{\includegraphics{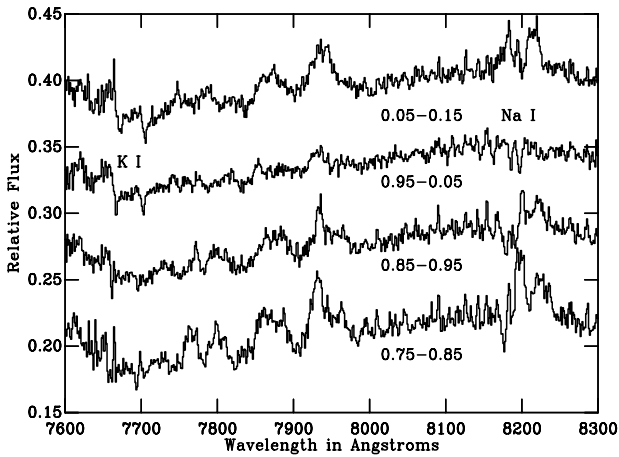}}
\vspace{2.5in}
\caption{ 
Phase-binned spectra of IP Peg from 1995 October 4th \& 5th. 
Phase ranges are marked, as are the K{\sc i} and Na{\sc i} absorption 
features. 
}
\label{d}
\end{figure}

The emission features are harder to identify, so to aid this
process we show the sums of spectra in four phase ranges near phase
zero in Fig. \ref{d}.
In this, and Fig. \ref{spectra} we can see emission features around
7770\AA, 7800\AA, 7870\AA, 7930\AA\ and 8200\AA.
These features are all broad, and weaken or disappear when the disc
is eclipsed in the phase 0.95-0.05 spectrum of Fig. \ref{d}.
This suggests a disc origin for all these lines.
The easiest to identify is the emission around 8200\AA, which is
where the Paschen series converges, but we can also expect disc
emission from the Na{I} doublet, and perhaps CaII 8203\AA\ and maybe even
HeII 8237\AA.
Friend et al. (1988) identify the OI triplet at
$\lambda\lambda$7771.9, 7774.2 and 7775.4\AA, which co-incides with
our 7770\AA\ feature.
This leaves three features still unexplained, of which only the
7800\AA\ feature appears in M89.

In principle there is a contribution to this spectrum from the F-star
used for telluric correction.
However, as stated in Section \ref{telluric} the strongest features
are around 10 percent of the continuum level.
As the summed spectrum is velocity shifted, these features are smeared
out over 600km s$^{-1}$, or 15 pixels, making them less than 1 percent
in depth.
The lines will be similarly smeared 
in the template spectra used for the cross-correlation, which is presumably
why they have no significant effect on the radial velocity study.


\section{A new ephemeris}
\label{ephemeris}

\begin{figure}
{\includegraphics{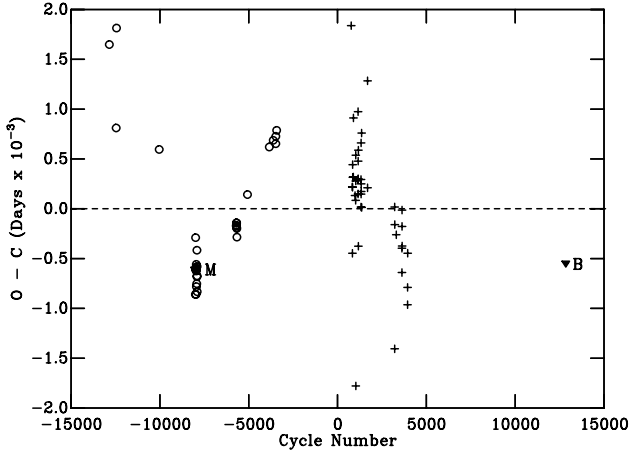}}
\vspace{2.5in}
\caption{
The O-C plot for the mid-white dwarf egress ephemeris from the
high-speed photometry timings of Wood et al. 1989 (circles), Wolf et 
al. 1993 (crosses) and the spectroscopic points from M89 
and this paper (filled triangles, also marked with M and B for clarity).
}
\label{o-c}
\end{figure}

Our spectroscopically derived eclipse timing-point and that of M89, both 
corrected to time of mid-white dwarf egress, were added to the 
photometric timings of Wood et al. (1989) and those of Wolf et al. (1993). 
All timings were converted to TDB. 
A linear ephemeris was then calculated to give
\[
\rm
JD(TDB) \;\; T_{egress} = 244\;7965.891\;44(5) + E*0.158\;206\;17(3)
\]
which is for the time of mid-white dwarf {\it egress}. Fig. \ref{o-c}
shows the $O-C$ plot for this ephemeris. 
We note in passing that the $O-C$ residuals calculated
from the linear ephemeris can be fitted by a sine wave with parameters
close to those given in Wolf et al. (1993), which they interpreted as a 
third body in the system.

\section{The apparent eccentricity}
\label{circularity}

\subsection{Are the new data consistent with a circular orbit?}

M89 found that an elliptical orbit was a better fit to their radial velocity 
data than a circular one. 
Here, we test our spectra for evidence of eccentricity.

We fitted each red-star radial-velocity data set with a circular orbit of 
the form
\[
  V = \gamma + K_{2}\sin(2\pi[\phi_{i}-\phi_{0}])
\]
and then an elliptical orbit of the form (e.g. Smart 1977, Chap. {\sc XIV};
Petrie 1962)
\[
  V = \gamma + K_{2}[\cos(\nu + \omega) + e\cos(\omega)]
\]
recording the mean square deviations ($S^{2}$) of the data from the fitted 
model in each case. In the elliptical orbit $\nu$ is the true anomaly of the 
red star, $\omega$ is the longitude of periastron of the red star from the 
centre of gravity of the system, and $e$ is the eccentricity of the red-star 
orbit. Following Lucy and Sweeney (1971), we use the $F$-test 
\begin{equation}
F = \frac{N_{\rm obs}-f_{\rm ell}}{f_{\rm ell}-f_{\rm crc}} 
    \left (\frac{S^{2}_{\rm crc}}{S^{2}_{\rm ell}} - 1 \right )
\end{equation}
to determine the significance of the elliptical fit over the circular one.
$N_{\rm obs}$ is the number of observations, $f_{\rm ell}$ is the number of 
degrees of
freedom in the elliptical fit, $f_{\rm crc}$ those in the circular fit, and 
$S^{2}_{\rm crc}, S^{2}_{\rm ell}$ are the
variances of the data from the circular and elliptical fits respectively,
i.e. $ S^{2} = \Sigma (y_{data} - y_{fit})^{2}$. This test may
be derived from the relationship between the $F$- and $\chi^{2}$-
distributions (e.g. Bevington 1969, Ch. 10).  
For both the circular and elliptical fits the systemic velocity,
$\gamma$, was held fixed at zero  since our method of radial velocity 
extraction uses an orbital-average as a template; the orbital period was 
also held fixed. The number of free parameters in each fit is then 
$f_{\rm ell}=4$ ($K_{2}, e, \nu$ and $\omega$) and $f_{\rm crc}=2$ ($K_{2}$ 
and $\phi_{0}$).

In statistical terms we are testing the hypothesis that the radial
velocity curves are consistent with zero eccentricity. 
Table \ref{ecc} shows for each radial velocity data set the
probability ($p$) that this hypothesis is correct, and since $p$
never falls below 30 percent, we must accept the hypothesis.

\subsection{Are the new data consistent with the previously
measured eccentricity?}

Although our data are consistent with a circular orbit, it is still
possible that, were they noisy enough, they could also be
consistent with an eccentricity at the level measured by M89.
So we determined the maximum level of eccentricity that could be present in 
our data.

For this exercise we fitted simultaneously the velocities from both TiO
bands, but retained the separate data sets for each resolution.
For each data set we constructed an $e$-$\chi^{2}$ space by
freezing $e$ at set values but allowing all other parameters (except
orbital period, always held fixed) to run free. 
Imposing the condition that $\chi^{2}_{\nu}$=1 for the $e$=0 fit, defines 
the error for each velocity point. 
From the 90\% confidence interval for five
free parameters (Lampton, Margon \& Bowyer 1976), both data sets yield an 
upper limit to any eccentricity in our data of around 0.05.

Thus we conclude that our upper limit to the eccentricity is
smaller than M89's detection (0.089$\pm$0.020). 

\subsection{Is the difference in eccentricity due to a difference
in technique?}

There are two important differences between our measurements and
those of M89.
They used the Na{I} doublet, whilst we used TiO; and they cross
correlated against an M-star template, whilst we used a 
velocity-shifted mean of the data.
So, we performed a cross-correlation study using the M3.5V
template GL27B, over the Na{I} region used by M89.
For testing the significance of the eccentricity we used $f_{\rm ell}=5$ 
and $f_{\rm crc}=3$, since $\gamma$ is an extra free parameter in each case. 
The results are shown in Table \ref{ecc}, and show that this study also
returns zero eccentricity.

The only remaining difference in technique is that M89
used a slightly different formula for the radial velocity, which
breaks down for $e>0.1$.
Although this approximation should not create a problem, for completeness we 
also tested their table 2 data, labelled M89 in our Table \ref{ecc}.  
Our code finds a similar eccentricity ($e=0.094$), rejecting a 
circular orbit at greater than 99.99\% confidence.

\subsection{Orbital eccentricity; a discussion}
\label{ecc_dis}

The conclusion is thus inescapable, the apparent eccentricity of the 
orbit has changed between the observations of M89 and our own. 
IP Peg is not the only system where an apparent eccentricity has been observed
in the radial velocity data. AM Her, U Gem, CH UMa and YY Dra all show
significant eccentricities (Friend et al. 1990a,b). 
The explanation that has been put forward to explain these deviations is 
irradiation of the secondary star, either by the white dwarf, bright spot
or a combination of both. 
Under this assumption, Davey \& Smith (1992) were
able to map the distribution of the Na{I} doublet across the surface of
the secondary star.
As Davey \& Smith point out, there is no satisfactory
theoretical explanation for the asymmetric distributions this model
predicts, nor is there any apparent correlation with other parameters which
would indicate the presence of strong heating.
As our spectra show the Na{I} doublet to be strongly contaminated by 
emission from the accretion disc, we put forward an alternative hypothesis, 
that this contamination, even when it is much weaker, affects the radial 
velocity curves of CVs.

IP Peg is a stunning example
of the random nature of deriving an eccentricity from radial velocity data:
it has been observed twice, over the same wavelength region, both times in
quiescence (although the M89 data may have been on the decline
from a small outburst), both use a similar procedure, both return the same 
value for the radial velocity of
the secondary star and yet one shows an eccentricity while the other does not.
The raison d'\^{e}tre for M89's correction to the radial velocity was the
apparent eccentricity of the orbit.
Significant irradiation would introduce an eccentricity (see M89), so the 
the absence of such effects (see also Section \ref{flux_deficits}), 
means that we should not make any correction to our semi-amplitude.


\section{Component masses}
\label{masses}

The mass function
\begin{equation}
 f(M) = \frac{P K_{2}^{3}}{2\pi G},
\end{equation}
where the symbols have their usual meaning, links the dynamical motion
of the secondary star to the mass of the unseen component. Using the
values obtained in Section \ref{rad_vel}, we find
\begin{equation}
 f(M) = 0.596 \pm 0.031\;\;M_{\odot}
\end{equation}
which acts as a lower limit to the mass ($M_{1}$) of the white dwarf star in 
IP Peg. Using this value, we can calculate masses of the individual 
components if the mass ratio, $q=M_{2}/M_{1}$ (where $M_{1}$ is the mass of
the white dwarf), and inclination, $i$, are known using
\begin{equation}
  M_{1} = (1+q)^{2} \frac {f(M)}{\sin^{3}(i)};\;\;\;\;\; M_{2} = q M_{1}.
\label{M1M2}
\end{equation}
The mass ratio may be determined by combining our $K_2$ with the 
rotational broadening measurement of Catal\'{a}n, Smith \& Jones (2000).
They find $V_{rot}\sin(i)=$125$\pm$15 km s$^{-1}$ where
\begin{equation}
 \frac{V_{rot}\sin(i)}{K_{2}} = (1 + q) f(q)
\end{equation}
and $f(q)$ is any of the approximations to the fractional Roche-lobe
radius $R_{2}/a$. 
Here we use the approximation due to Eggleton (1983), which gives a mass 
ratio of $q=0.322\pm 0.075$. 
We derive the inclination using the eclipse width of $\Delta\phi=0.0863$ 
from Wood \& Crawford (1986).
To be consistent with this eclipse width, we must raise our lower limit on $q$
to 0.285, since smaller mass ratios will not produce a long enough eclipse,
even if $i=90$ degrees.
This sets the upper bound of the inclination to 90 degrees, and
thus our measurement of the inclination is $i=85.9^{+4.1}_{-2.9}$ degrees.
Since the error in $V_{rot}\sin(i)$ dominates the errors for $q$, $i$,
$M_1$ and $M_2$, we can simply use the extreme values of $i$, with their
corresponding values of $q$ in equation \ref{M1M2} to obtain our limits
on the component masses.  
These are 
$M_{1}=1.05^{+0.14}_{-0.07}$M$_{\odot}$ and 
$M_{2}=0.33^{+0.14}_{-0.05}$M$_{\odot}$.

Martin et al. (1987) find, from the reduction
in cross-correlation errors, that the secondary star is probably closer to 
M4.5V than M3.5V. 
From the TiO line-depth ratio, Catal\'{a}n et al (2000) find the
spectral type to be M5$\pm$0.5V. 
Using the observationally derived mass-spectral type relation of 
Kirkpatrick \& McCarthy (1994--their equation (11)), which is 
supported by the theoretical models of Baraffe \& Chabrier (1996), our value 
of $M_{2}=0.35 \pm 0.08$M$_{\odot}$ implies a spectral type of 
M2.2$\pm$0.6V. 
The error in the calculated spectral type includes both the error 
in our mass estimate and the error in the relation itself. For a spectral
type of M4V, this same relation returns a mass of 0.18M$_{\odot}$. 
Although our mass estimate is now consistent with an M-dwarf, the 
secondary star in IP Peg is still over-massive for its derived spectral
type.

We can also calculate the radius for the secondary star, and for consistency
do so using equation (3) of Smith \& Dhillon (1998).
This gives a radius of $R_{2}=0.40 \pm 0.03$R$_{\odot}$. 
We can then place the star in the mass-radius-period relationships
given in that paper.
In all cases IP Peg is now consistent not only with the other CVs, but also
with the current semi-empirical and theoretical relationships.

Our value for $q$ is low compared with previous determinations, and although 
the error bars overlap with those of Wood \& Crawford (1986) who find 
$ 0.35 < q < 0.49 $ from measurements of the bright spot, it is probably 
inconsistent with that of Wolf et al (1993) who measure $q=0.6$ (no error 
given) from a similar technique.
Both values rely on the center-of-light for the stream following a ballistic
trajectory, which may well not be the case once it begins to interact with
the Keplerian flow in the disc.
Marsh (1988) also find $q=0.6$, but his value relies on measuring the
radial velocity of the primary using the wings of emission lines from 
the disc.
Long \& Gilliland (1999) have shown how this technique is inaccurate for
U Gem.


\section{The secondary star lightcurve} 
\label{flux_deficits}

Measuring the flux in the TiO bands should allow us to extract the flux
of the secondary star alone, assuming there is no contribution to the
TiO from the other sources of the light in the system.
Wade \& Horne (1988) constructed such a lightcurve for the short period
dwarf nova Z Cha, by measuring the total flux in the TiO band with
respect to a fitted continuum ({\it i.e.} a flux deficit).
Here we undertake a similar study, but using a slightly more sophisticated
method.

\subsection{The method}

If the temperature dependence of the equivalent width of the spectral
features is small for the range of temperatures over the secondary star, 
the flux from the star at any given phase is some phase-averaged 
secondary-star spectrum multiplied by a scaling factor we refer to as 
$k_\phi$.
Thus, if we scale and then subtract the phase-averaged spectrum from the
spectrum at a given phase, the residual will be smooth across the TiO bands
when we have chosen the right $k_\phi$.
Since both the phase-averaged spectrum, and the spectrum at the phase under
consideration have contributions from the accretion disc, the residual is
not simply the accretion disc flux at the phase in question.
However, if the disc has a spectrum which is approximately
a straight line at all phases, the residual will also be a straight line.

To determine $k_\phi$ for a given spectrum we subtract from it varying
portions of the mean spectrum, fit the residual with a straight line,
and evaluate $\chi^2$ for the fit.
We then look for the value of $k_\phi$ which minimises $\chi^2$.
To minimise the effect of the assumption of a straight line fit, we
evaluate $\chi^2$ over a relatively short fragment of spectrum.

\begin{figure}
{\includegraphics{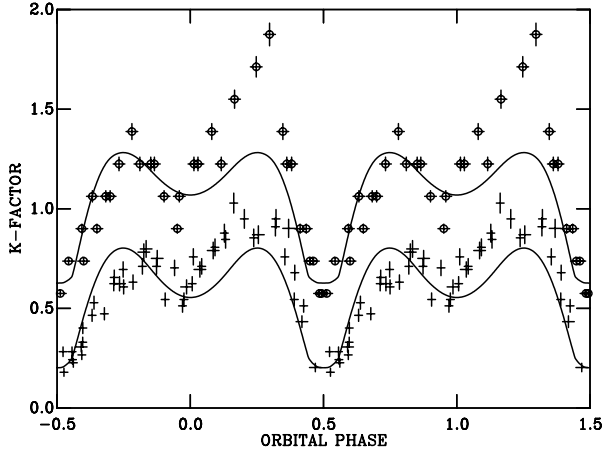}}
\vspace{2.5in}
\caption{ 
The $k_{\phi}$-factor plotted against phase for 
the lower resolution data ( Oct. 4th \& 5th, upper curve, displaced
upwards by 0.25 for clarity) and for the higher resolution data (Oct.
9th \& 10th, lower curve, displaced downwards by 0.4). 
The region fitted was 7450--7750\AA. 
Shown over-plotted (full line) is the best-fitting `ellipsoidal modulation 
with eclipses' model. 
The data are plotted twice and $k_{\phi}$ is normalised to be about one
at phase zero.
}
\label{kfac_long}
\end{figure}

\begin{figure}
{\includegraphics{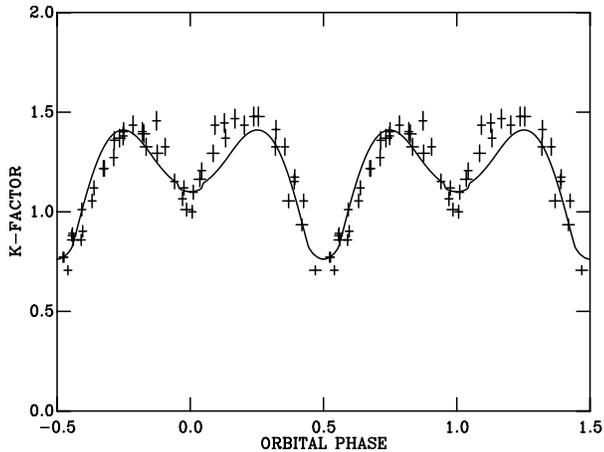}}
\vspace{2.5in}
\caption{
As the previous figure but derived fitting the region 7040--7340\AA\ in
the higher resolution data.
}
\label{kfac_short}
\end{figure}

\subsection{Results}

Fig. \ref{kfac_long} shows $k_{\phi}$ plotted against orbital phase
calculated from the region between 7450--7750\AA\ for both the 
high and low resolution data sets. 
The average $(\chi_{\nu}^{2})$s (over all spectra) for the individual 
straight-line fits to the residuals were 2.21$\pm$0.23 and 3.59$\pm$0.45 
for the high and low resolution data, respectively (the error is the standard 
deviation of the individual fits around these means).
Fig. \ref{kfac_short} shows $k_{\phi}$ plotted 
against orbital phase for the case of a straight line fit to the region 
between 7040--7340\AA\ for the low resolution data set.

The most prominent feature in Figs. \ref{kfac_long} and \ref{kfac_short}
is the dip at phase 0.5.
The obvious interpretation for this is that it is an eclipse of the 
secondary star by the accretion disc.
This is a rather remarkable result, since the accretion discs of quiescent
dwarf novae are normally thought to be optically thin.
However, we are viewing the disc in IP Peg at an unusual angle (nearly edge
on), and whilst it may be that the discs are optically thin when viewed from
the pole, the large column length to the secondary star at phase
0.5 may be enough to obscure it.
Alternatively, it may be that there are some optically thick regions in the
disc, which cause the eclipse we see.

At other phases the lightcurves are broadly consistent with the 
double-humped ellipsoidal modulation we might expect.
To see just how close they are to this expectation, and prove that the 
suggestion of an eclipse is at least plausible, we modelled the data
using the code described in Ioannou et al (1999). 
The model is a standard ellipsoidal model with Roche geometry, but
importantly for this work, includes the mutual eclipses between the 
disc and secondary star. 
Here the disc is assumed to be a cold (i.e. dark) obscurer.
We used the parameters derived in Section \ref{masses} in addition to
a limb darkening coefficient of 0.4, a gravity darkening exponent of 0.08 
and a secondary star (pole) temperature of 3375K.
We performed a grid search in opening angle and radius to find the best
fit to the data.

In Figs. \ref{kfac_long} and \ref{kfac_short} we plot the best fitting
models.
This shows that the model can account for the broad outline of the 
lightcurves, but there are problems in detail.
This is reflected by the fact that the best $\chi_{\nu}^2$ obtained
was 6.1 (for the data of Fig \ref{kfac_short}), and that all the datasets
required a disc radius equal to the Roche-lobe radius of the primary (the 
maximum we allowed).
For the lower resolution data of Fig. \ref{kfac_long} the best fit opening
angle was 8$^{\circ}$, 4$^{\circ}$ for the higher resolution.
For Fig. \ref{kfac_short} the opening angle was 3$^{\circ}$. 

The models throw into sharp relief how the flux at $\phi=0.75$ is less  
than that at $\phi=0.25$.
Furthermore, Fig. \ref{kfac_long} shows how this asymmetry can vary from
night to night.
Similar night-to-night variability was seen in the Z Cha TiO flux deficits
by Wade \& Horne (1988). 
The asymmetry, and its variability, could be explained by strong irradiation 
from the bright spot depleting the
TiO on the side of the red star facing the bright spot, i.e. the side seen
at $\phi=0.75$.
However, this would produce a marked eccentricity in the
radial velocity curves deduced from the TiO regions of the spectrum. 
As shown in Section \ref{circularity}, no significant eccentricity
is found from any of the IP Peg absorption features. 

The models suggest that the problem is with the data around phase 0.25.
This means obscuration of the secondary star by the disc is unlikely, given
the star's position well clear of the disc at this phase.
Thus we suspect the problem lies with our simple contamination model (a
straight line fit).
The spectral variations of the disc with time are probably more complex.


\section{Conclusions}

The main conclusions from this work are as follows.
\hfil\break
(1) There is some form of variability which means that radial velocity 
studies of the infrared Na{I} doublet in IP Peg can sometimes return 
an apparently elliptical orbit.
We suspect this is contamination from disc emission.
\hfil\break
(2) We strongly recommend the use of TiO rather than Na{I} for future 
radial velocity studies.
Not only does it avoid the contamination problem described above, but
it gives a factor two gain in signal-to-noise.
\hfil\break
(3) Our TiO study suggests the radial velocity semi-amplitude for the secondary
star should be revised to $K_{2}=331.3\pm 5.8$km s$^{-1}$, and as a result
the primary and secondary star masses become 
$M_{1}=1.05^{+0.14}_{-0.07}$M$_{\odot}$ and 
$M_{2}=0.33^{+0.14}_{-0.05}$M$_{\odot}$
respectively.
\hfil\break
(4) Despite this downwards revision of the secondary star mass, 
it is still over-massive for its observed spectral type, but now agrees
with the current semi-empirical and theoretical mass-radius-period
relationships.
\hfil\break
(5) The accretion disc eclipses the secondary star in quiescence, implying
that, when viewed at high inclination, it is optically thick.


\section*{Acknowledgements}

The INT and JKT is operated on the island of La Palma by the Isaac Newton 
Group in the Spanish Observatorio del Roque de los Muchachos of the 
Instituto de Astrofisica de Canarias. Data reduction was carried out on 
the Keele Starlink node using the {\sc ark} software. 
Data on the outburst state of IP Peg  
was generously provided by the AAVSO International Database and we gratefully
acknowledge with thanks all the variable star observers worldwide who 
participate in this program. We also thank Sandi Catal\'{a}n, Janet Wood and 
Rob Jeffries for useful discussions.



\end{document}